\documentclass[12pt]{iopart}
\usepackage{graphicx}

\begin{document}

\title[]{Phase diagram, band structure and density of states in two-dimensional attractive Fermi-Hubbard model with Rashba spin-orbit coupling}

\author{Rui Han, Feng Yuan, Huaisong Zhao \footnote[1] {Author to whom any correspondence should be addressed.}}
\address{College of Physics, Qingdao University, Qingdao 266071, People's Republic of China}
\ead{hszhao@qdu.edu.cn}
\vspace{10pt}
\begin{indented}
\item[]May 2022
\end{indented}

\begin{abstract}
Based on the two-dimensional (2D) attractive Fermi-Hubbard model with Rashba spin-orbit coupling (SOC), the SOC strength and Zeeman field dependences of the phase diagram are investigated by calculating the pairing gap self-consistently. The results reveal that the phase transition from the BCS superfluid to the topological superfluid happens under proper Zeeman field strength and SOC strength. In particular, in contrast to the BCS superfluid decreasing monotonically as the SOC strength increasing, the topological superfluid region  shows a dome  with the SOC strength increasing. An optimal region in the phase diagram to find the topological superfluid can be found, which is important to realize the topological superfluid in optical lattice experimentally. Then we obtain the change of both band structure and density of states (DOS) during the topological phase transition, and explain the four peaks of DOS in the topological superfluid by the topology change of the low-energy branch of quasiparticle energy spectra. Moreover, the topological superfluid can be suppressed by the doping concentration.

\end{abstract}

%
%
\submitto{}
%
%
%

\section{Introduction}
Rashba spin-orbit coupling (SOC) is closely related to many non-trivial physical properties \cite{Zhai15,Jayantha11,Jayantha11-2,Hu2011,He2013,Vyasanakere12,Zhang2013}, in particular for the topological superfluid state\cite{Zhou11,Liao10,Cao14,Zhou14,Hu13}. The topological superfluid state in the ultracold Fermi atomic system had been proposed by the coexistence of Rashba SOC, an external Zeeman field and s-wave pair order \cite{Alicea2012}. However, the Rashba SOC has not yet been realized experimentally until now.  In recent years, by engineering the atom-laser interaction, the synthetic magnetic field, synthetic electric field and synthetic low-dimensional Raman-type SOC in the ultracold neutral atoms can be realized \cite{Lin2009,Lin2011,Lin2011-2,Galitski13,Cheuk12,Wang12,Zhang14,Williams13,Meng16,Huang18,Burdick16,Huang16,Wu2016,Zhang18}. For the Raman-type SOC system, theoretically the topological superfluid  will appear in one-dimensional (1D) Fermi atomic system, but it does not exist in two-dimensional (2D) Raman-type case.
 Degenerate $^{40}$K Fermi gases with 2D Raman-type SOC had been realized with a perpendicular Zeeman field \cite{Meng16} or an in-plane Zeeman field \cite{Huang16}, respectively. To find the topological superfluid, we focus on how to realize the Rashba SOC in 2D ultracold atoms. Since the experimental breakthroughs in realization of Raman-type SOC in ultracold atomic gases, the region of topological superfluid state in phase diagram receives significant attention, which plays an important role in finding the topological superfluid state in experiments. Theoretically the phase diagrams were investigated  in 2D polarized Fermi gas with Rashba SOC \cite{Zhou11,Cao14,Yang2012,Liu2012,Sheehy2007,Zhang13}. J. Zhou and coworkers systematically studied the  phase separation and gave the optimal parameter region for the preparation of the topologically nontrivial superfluid phase across a BCS-BEC crossover \cite{Zhou11}.

The periodic optical lattice potential confining cold atoms can be generated by superimposing orthogonal standing waves \cite{Bloch08}. Experimentally the attractive Fermi-Hubbard model in cold atom systems had been realized \cite{Mitra18,Hackermuller10,Peter20}, and the quantum phase transitions were discussed well \cite{Scalettar89,Moreo91,Singer96,Paiva04,Ho04}. In particular, the optical lattice with the Rashba SOC  is a good platform to study the topological phase transition \cite{Xu2014,Koinov17,Sun2013,Gremaud13,Riera13,Goldman2009,Iskin2013,Burrello2013,Wang2013,Qu2013,Jia2019,Wu2021}. By considering an in-plane Zeeman field in Fermi-Hubbard model, mean-field phase diagrams of 2D square optical lattice with Rashba SOC were given \cite{Xu2014}, and both Fulde-Ferrell (FF) and Larkin-Ovchinnikov (LO) phases  may be observed \cite{Liu07,Wu13,Liu2013,Yuan2021}. A semimetal-superfluid transition in an optical lattice with the on-site attraction was induced by the SOC \cite{Sun2013}.  N. Goldman and coworkers found that when the system enters the non-Abelian gauge field regime, the two Van Hove singularities (VHS) in the density of states (DOS) of an ultracold Fermi gas are split into four \cite{Goldman2009}, which leads to that there are four peaks of the DOS in the non-Abelian gauge field regime. Based on a self-consistent mean-field theory in the optical lattice with the repulsive on-site interaction,  L. Wang and coworkers discussed a quantum phase transition from a semimetallic phase to a band insulator and found the change of the energy spectra and DOS during the phase transition \cite{Wang2013}.

Although the Rashba-type SOC has not yet been realized experimentally, several theoretical proposals on how to realize it had been given \cite{Wang2018,Juzeliunas10,Campbell11,Xu12,Xu13,Zhou19,Sau11,Anderson13,Dalibard11}. And the optimal parameters to find the topological superfluid state in the Rashba SOC are still an open question.
In this work, based on the mean-field theory of Fermi atom system, we study the phase transition from BCS superfluid to topological superfluid in 2D attractive Fermi-Hubbard model with synthetic Rashba spin-orbit coupling. Our results show that the optimal topological superfluid region exists when both the Zeeman field strength and Rashba SOC strength take appropriate values, i.e., the Zeeman field strength has a large adjustable range and the pairing gap is still large. Moreover, we discuss the effect of Rashba SOC interaction on the band structure and DOS during the phase transition
from BCS superfluid to topological superfluid, then find the characteristic signals to distinguish two superfluid states. Although the quantum fluctuation of 2D system is large, the mean-field theory can still be expected to give a qualitatively reliable prediction.

Our paper is organized as follows. In the sections 2 and 3, we introduce the Green's functions and the phase diagrams, respectively, then discuss the angle resolved photoemission spectroscopy (ARPES) and band structure in section 4. The SOC strength dependence of  DOS is studied in section 5. Moreover, we discuss the doping dependent of phase diagrams in section 6, and give a summary in section 7.
\section{Mean-field description with Green's functions}
It is widely believed that the attractive Fermi-Hubbard model captures the main physical properties of the Fermi superfluid in optical lattices. Therefore, it is important to study the attractive Fermi-Hubbard model with the SOC and Zeeman fields.
For two-dimensional optical lattice with Rashba SOC and a perpendicular Zeeman field, the Hamiltonian can be described by the Fermi-Hubbard model,
\begin{eqnarray}\label{Humodel2}
 H&=&-t\sum_{<ij>}C_{i\sigma}^{\dagger}C_{j\sigma}-U\sum_{i}C_{i\uparrow}^{\dagger}C_{i\downarrow}^{\dagger}C_{i\downarrow}C_{i\uparrow}
 -(\mu+h\sigma_{z})\sum_{i\sigma}C_{i\sigma}^{\dagger}C_{i\sigma}\nonumber\\
 &+&\lambda\sum_{i}[C_{i+x\downarrow}^{\dagger}C_{i\uparrow}-C_{i+x\uparrow}^{\dagger}C_{i\downarrow}
 +i(C_{i+y\downarrow}^{\dagger}C_{i\uparrow}+C_{i+y\uparrow}^{\dagger}C_{i\downarrow})]+H. c.,
\end{eqnarray}
where $<ij>$ means the nearest-neighbor sites of 2D lattice, $\mu$ is the chemical potential, and $U>0$ is the on-site attraction strength. Parameters $t$, $\lambda$ and $h$ are strength of hopping, Rashba SOC and Zeeman field, respectively. $C_{i\sigma}^{\dagger}(C_{i\sigma})$ is the atom creation (annihilation) operator with spin $\sigma$ ($\sigma=\uparrow,\downarrow$) at the lattice $i$. Here $\sigma_{z}$ is the Pauli matrix. Within the mean field theory, the interaction part with the four operators can be dealt with $U\sum_{i}C_{i\uparrow}^{\dagger}C_{i\downarrow}^{\dagger}C_{i\downarrow}C_{i\uparrow}=\sum_{i}(\Delta^{*}C_{i\downarrow}C_{i\uparrow}+\Delta C_{i\uparrow}^{\dagger}C_{i\downarrow}^{\dagger})$,
where the pairing gap $\Delta=U<C_{i\downarrow}C_{i\uparrow}>$.
 Equation (\ref{Humodel2}) has a simple BCS form within the mean-field theory in momentum space,
\begin{eqnarray}\label{tjbm}
H&=&\sum_{{\bf k},\sigma}(\xi_{\bf k}-h\sigma_{z})C^{\dagger}_{{\bf k}\sigma}C_{{\bf k}\sigma}+\sum_{\bf k}[\lambda_{\rm so}({\bf k})C^{\dagger}_{{\bf k}\uparrow}C_{{\bf k}\downarrow}+\lambda^{*}_{\rm so}({\bf k})C^{\dagger}_{{\bf k}\downarrow}C_{{\bf k}\uparrow}]\nonumber\\
&-&\sum_{{\bf k}}(\Delta^{*}C_{{\bf k}\downarrow}C_{-{\bf k}\uparrow}+\Delta C^{\dagger}_{-{\bf k}\uparrow}C^{\dagger}_{{\bf k}\downarrow}),
\end{eqnarray}
where $\xi_{\bf k}=-Zt\gamma_{\bf k}-\mu$, $\lambda_{\rm so}({\bf k})=\lambda(\sin{k_{x}}+i\sin{k_{y}})$, $\lambda^{*}_{\rm so}({\bf k})=\lambda(\sin{k_{x}}-i\sin{k_{y}})$, $\gamma_{\bf k}=0.5(\cos{k_{x}}+\cos{k_{y}})$, and the parameters $Z=4$ for 2D square lattice. The Hubbard energy $U$ and lattice constant $a_{0}=1$ can be used as the units of energy and length. In this study, $t/U=0.3$, temperature $T/U=0.001$.
Then we define the diagonal Green's function with spin up $G_{1}({\bf k},\tau-\tau^{'})=-\langle T C_{{\bf k}\uparrow}(\tau)C^{\dagger}_{{\bf k}\uparrow}(\tau^{'})\rangle$  and spin down $G_{2}({\bf k},\tau-\tau^{'})=-\langle T C_{{\bf k}\downarrow}(\tau)C^{\dagger}_{{\bf k}\downarrow}(\tau^{'})\rangle$.
Then we define the off-diagonal Green's function with pairing related terms $\Gamma^{\dagger}({\bf k},\tau-\tau^{'})=-\langle T C^{\dagger}_{{\bf -k}\downarrow}(\tau)C^{\dagger}_{{\bf k}\uparrow}(\tau^{'})\rangle$.
The two SOC related terms, $S({\bf k},\tau-\tau^{'})=-\langle T C_{{\bf k}\downarrow}(\tau)C^{\dagger}_{{\bf k}\uparrow}(\tau^{'})\rangle$,
$F^{\dagger}({\bf k},\tau-\tau^{'})=-\langle T C^{\dagger}_{{\bf -k}\uparrow}(\tau)C^{\dagger}_{{\bf k}\uparrow}(\tau^{'})\rangle$.
 Based on the motion equation of Green's function \cite{Zhao20}, we can obtain these functions,
\begin{eqnarray}\label{Greenfunction}
G_{1}({\bf k},\omega)=\sum_{a}\left[\frac{U'^2_{a{\bf k}}}{\omega-E_{a{\bf k}}}+
\frac{V'^2_{a{\bf k}}}{\omega+E_{a{\bf k}}}\right],
G_{2}({\bf k},\omega)=\sum_{a}\left[\frac{U^2_{a{\bf k}}}{\omega-E_{a{\bf k}}}+
\frac{V^2_{a{\bf k}}}{\omega+E_{a{\bf k}}}\right],\nonumber\\
\Gamma^{\dagger}({\bf k},\omega)=\sum_{a}\left[
\frac{\alpha_{a{\bf k}}}{\omega+E_{a{\bf k}}}+\frac{\beta_{a{\bf k}}}{\omega-E_{a{\bf k}}}\right],
S({\bf k},\omega)=\sum_{a}\left[\frac{\lambda^{*}_{\rm so}({\bf k})P_{a{\bf k}}}{\omega-E_{a{\bf k}}}+
\frac{\lambda^{*}_{\rm so}({\bf k})Q_{a{\bf k}}}{\omega+E_{a{\bf k}}}\right],\nonumber\\
F^{\dagger}({\bf k},\omega)=\sum_{a}\left[\frac{\lambda^{*}_{\rm so}({\bf k})T'_{a{\bf k}}}{\omega-E_{a{\bf k}}}+
\frac{\lambda^{*}_{\rm so}({\bf k})W'_{a{\bf k}}}{\omega+E_{a{\bf k}}}\right],
\end{eqnarray}
where $U'^2_{a{\bf k}}=(\Omega'_{a{\bf k}}+\Xi'_{a{\bf k}})/\Sigma_{a{\bf k}}$, $V'^2_{a{\bf k}}=(\Omega'_{a{\bf k}}-\Xi'_{a{\bf k}})/\Sigma_{a{\bf k}}$, $U^2_{a{\bf k}}=(\Omega_{a{\bf k}}+\Xi_{a{\bf k}})/\Sigma_{a{\bf k}}$, $V^2_{a{\bf k}}=(\Omega_{a{\bf k}}-\Xi_{a{\bf k}})/\Sigma_{a{\bf k}}$, $\alpha_{a{\bf k}}=[E^{2}_{a{\bf k}}+2hE_{a{\bf k}}+h^{2}-\xi^{2}_{\bf k}-\Delta^{2}-\lambda^{2}_{\rm so}({\bf k})]\Delta/\Sigma_{a{\bf k}}$, $\beta_{a{\bf k}}=-[E^{2}_{a{\bf k}}-2hE_{a{\bf k}}+h^{2}-\xi^{2}_{\bf k}-\Delta^{2}-\lambda^{2}_{\rm so}({\bf k})]\Delta/\Sigma_{a{\bf k}}$, $P_{a{\bf k}}=[E^{2}_{a{\bf k}}+2\xi_{\bf k}E_{a{\bf k}}+\xi^{2}_{\bf k}-h^{2}-\Delta^{2}-\lambda^{2}_{\rm so}({\bf k})]/\Sigma_{a{\bf k}}$, $Q_{a{\bf k}}=-[E^{2}_{a{\bf k}}-2\xi_{\bf k}E_{a{\bf k}}+\xi^{2}_{\bf k}-h^{2}-\Delta^{2}-\lambda^{2}_{\rm so}({\bf k})]/\Sigma_{a{\bf k}}$, $T'_{a{\bf k}}=2(h+\xi_{\bf k})\Delta/\Sigma_{a{\bf k}}$,
 $W'_{a{\bf k}}=-T'_{a{\bf k}}$,
$\Sigma_{a{\bf k}}=2(E^{2}_{a{\bf k}}-E^{2}_{a'{\bf k}})E_{a{\bf k}}$, $\Omega'_{a{\bf k}}=E^{3}_{a{\bf k}}-[(\xi_{\bf k}+h)^{2}+\Delta^{2}+\lambda^{2}_{\rm so}({\bf k})]E_{a{\bf k}}$,
$\Xi'_{a{\bf k}}=(\xi_{\bf k}-h)E^{2}_{a{\bf k}}-(\xi_{\bf k}+h)[\xi^{2}_{\bf k}+\Delta^{2}-h^{2}-\lambda^{2}_{\rm so}({\bf k})]$,
$\Omega_{a{\bf k}}=E^{3}_{a{\bf k}}-[(\xi_{\bf k}-h)^{2}+\Delta^{2}+\lambda^{2}_{\rm so}({\bf k})]E_{a{\bf k}}$,
$\Xi_{a{\bf k}}=(\xi_{\bf k}+h)E^{2}_{a{\bf k}}-(\xi_{\bf k}-h)[\xi^{2}_{\bf k}+\Delta^{2}-h^{2}-\lambda^{2}_{\rm so}({\bf k})]$, $a(a')=1,2$, and $a'\neq a$. The weight factors $U^2_{a{\bf k}} (U'^2_{a{\bf k}})$, $V^2_{a{\bf k}} (V'^2_{a{\bf k}})$ satisfy the sum rule: $\sum_{a}(U^2_{a{\bf k}}+V^2_{a{\bf k}})=1$,  $[\sum_{a}(U'^2_{a{\bf k}}+V'^2_{a{\bf k}})=1]$.
And the quasiparticle energy spectra
\begin{eqnarray}\label{Quasi-en-spec}
E_{1\bf{k}}&=&\sqrt{h^{2}+\xi^{2}_{\bf k}+\Delta^{2}+\lambda^{2}_{\rm so}({\bf k})+2\sqrt{h^{2}(\xi^{2}_{\bf k}+\Delta^{2})+\xi^{2}_{\bf k}\lambda^{2}_{\rm so}({\bf k})}}\nonumber\\
E_{2\bf{k}}&=&\sqrt{h^{2}+\xi^{2}_{\bf k}+\Delta^{2}+\lambda^{2}_{\rm so}({\bf k})-2\sqrt{h^{2}(\xi^{2}_{\bf k}+\Delta^{2})+\xi^{2}_{\bf k}\lambda^{2}_{\rm so}({\bf k})}}
\end{eqnarray}
The chemical potential $\mu$ and pairing gap $\Delta$ are determined by the equations $n=\sum_{\sigma}n_{\sigma}$ and $\Delta=U<C_{i\uparrow}^{\dagger}C^{\dagger}_{i\downarrow}>$, where $n$ is the average occupancy number per lattice site. At half-filling, $n=1$. Based on the diagonal and off-diagonal Green's functions, we can get the self-consistent equations  as,
\begin{eqnarray}\label{twoequtions}
n&=&\sum_{{\bf k,a}}[(U'^2_{a{\bf k}}+U^2_{a{\bf k}})n_{F}(E_{a\bf{k}})+(V'^2_{a{\bf k}}+V^2_{a{\bf k}})n_{F}(-E_{a\bf{k}})]\nonumber\\
 1&=&\sum_{{\bf k,a}}[\frac{\alpha_{a{\bf k}}-\beta_{a{\bf k}}}{2E_{{a\bf k}}}n_{F}(E_{a\bf{k}})+\frac{\beta_{a{\bf k}}}{2E_{{a\bf k}}}]
\end{eqnarray}
where $n_{F}(E_{a\bf{k}})$ are Fermi distributions. The parameters $\Delta$, $\mu$ are obtained without using adjustable parameters. In particular, at the topological phase transition point, we find the quasiparticle energy spectra $E_{{2\bf k_{M}}}=0$ at ${\bf k_{M}}=[\pm\pi,0],[0,\pm\pi]$ (excitation gap vanishes while the pairing gap remains finite), thus $U'^2_{a{\bf k_{M}}}$, $U^2_{a{\bf k_{M}}}$, $V'^2_{a{\bf k_{M}}}$, $V^2_{a{\bf k_{M}}}$, $\alpha_{a{\bf k_{M}}}$, and $\beta_{a{\bf k_{M}}}$ in the self-consistent equations above diverge at ${\bf k_{M}}$. However, in fact, these divergence points can be solved analytically. Therefore, the self-consistent equations in the topological transition point are expressed as,
\begin{eqnarray}\label{twoequtions-topological}
n&=&\sum_{{\bf k\ne k_{M},a}}[(U'^2_{a{\bf k}}+U^2_{a{\bf k}})n_{F}(E_{a\bf{k}})+(V'^2_{a{\bf k}}+V^2_{a{\bf k}})n_{F}(-E_{a\bf{k}})]+\sum_{{\bf k_{M}}}f({\bf k_{M}})\nonumber\\
 1&=&\sum_{{\bf k\ne k_{M},a}}[\frac{\alpha_{a{\bf k}}-\beta_{a{\bf k}}}{2E_{{a\bf k}}}n_{F}(E_{a\bf{k}})+\frac{\beta_{a{\bf k}}}{2E_{{a\bf k}}}]+\sum_{{\bf k_{M}}}g({\bf k_{M}})\nonumber\\
 E_{{2\bf k_{M}}}&=&0
\end{eqnarray}
where $f({\bf k_{M}})=(U'^2_{1{\bf k_{M}}}+U^2_{1{\bf k_{M}}})n_{F}(E_{1\bf{k_{M}}})+(V'^2_{1{\bf k_{M}}}+V^2_{1{\bf k_{M}}})n_{F}(-E_{1\bf{k_{M}}})+(\xi^{2}_{\bf k_{M}}+\Delta^{2}+h^{2}+\lambda^{2}_{\rm so}({\bf k_{M}}))/{E^{2}_{1{\bf k_{M}}}}$, $g({\bf k_{M}})=(\alpha_{1{\bf k_{M}}}-\beta_{1{\bf k_{M}}})n_{F}(E_{1\bf{k_{M}}})/{2E_{{1\bf k_{M}}}}+\beta_{1{\bf k_{M}}}/{2E_{{1\bf k_{M}}}}-{h}/{E^{2}_{1{\bf k_{M}}}}$ then the parameters in the topological transition point are obtained.

\section{Phase diagram in the $h-\lambda$ plane at half-filling}
\begin{figure}[ht]
  \centering
  \includegraphics[width=0.9\textwidth]{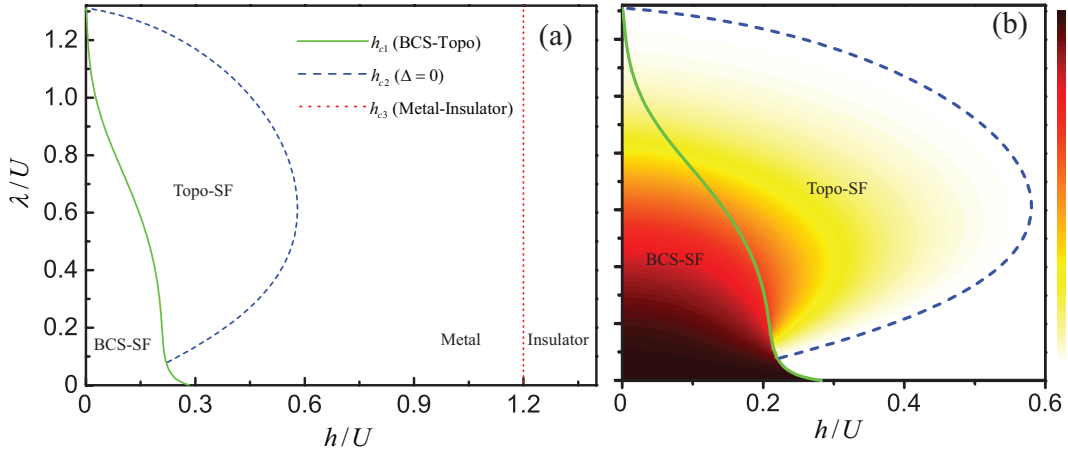}
  \vspace{-3mm}
  \caption{(a) Phase diagram in the $h-\lambda$ plane of 2D optical lattice with Rashba SOC effect (BCS-SF, Topo-SF, normal metal phase and insulator phase); (b) Pairing gap $\Delta$ as functions of $h$ and $\lambda$ for $t/U=0.3$, $n=1.0$, $T/U=0.001$. And $h_{c1}$(green solid line), $h_{c2}$ (blue dashed line) and $h_{c3}$ (red dotted line) mark the BCS-topological superfluid transition, superfluid-normal metal transition and normal metal-insulator transition, respectively. }
  \vspace{-3mm}
  \label{fig1} 
\end{figure}

 As $h$ increases, a transition from a nontopological BCS superfluid to a topological superfluid state appears at the critical Zeeman field strength $h_{c1}$. Then stronger Zeeman field makes the superfluid disappear and the normal metal phase appears, i.e., $\Delta=0$ at the superfluid transition point $h_{c2}$, which can be solved simultaneously with the state of equations (\ref{twoequtions}) to obtain $\mu$ and $h_{c2}$.  What's more, a metal-insulator transition appears at $h_{c3}$ owing to the open of band gap which is the result of the Zeeman splitting.  The phase diagrams in the $h-\lambda$ plane for half-filling $n=1.0$ are shown in figure \ref{fig1}. And $h_{c1}$ (green solid line), $h_{c2}$ (blue dashed line) and $h_{c3}$ (red dotted line) mark the BCS-topological superfluid transition, superfluid-normal metal transition and normal metal-insulator transition, respectively.

There are four distinct phase-separated regions in the phase diagram, namely, the BCS superfluid (BCS-SF), topological superfluid (Topo-SF), normal metal and insulator phases. With the increase of $\lambda$, the region of the BCS superfluid decreases monotonically and disappears at $\lambda/U=1.31$. However, in contrast to the case of the BCS superfluid, the topological superfluid region between green solid line and blue dashed line shows a dome as a function of $\lambda$, i.e., it is proportional to $\lambda$ at the low $\lambda$ region, and reaches a maximum at the optimal $\lambda_{op}/U=0.61$, then decreases when $\lambda>\lambda_{op}$.  Therefore, to realize the topological superfluid of the Rashba SOC in optical lattice experimentally, it is necessary to select a reasonable $\lambda$ and $h$. In particular, the $\lambda$ has a optimal region $\lambda_{op}/U=0.61$, where $h$ has a large adjustable range and the pairing gap is still large. The metal-insulator phase transition happens at $h_{c3}/U=1.2$, and related discussion had been given in 1D optical lattice with SOC \cite{Han2022}. In figure \ref{fig1}b, we can find an interesting phenomenon, i.e., the pairing gap $\Delta$ as a function of $h$ in the low $\lambda$ region decreases precipitously around the topological transition point $h_{c1}$ while it decreases gently with the increase of $h$ in the larger $\lambda$ region. Based on the different $h$-dependent behavior of parameters in different $\lambda$ region, first-order phase transition appears in the low $\lambda$ region and the second-order phase transition happens in the larger $\lambda$ region. Similar case had also been discussed in a trapped 2D polarized Fermi gas with Rashba SOC \cite{Zhou11}.

\begin{figure}[ht]
  \centering
  \includegraphics[width=0.9\textwidth]{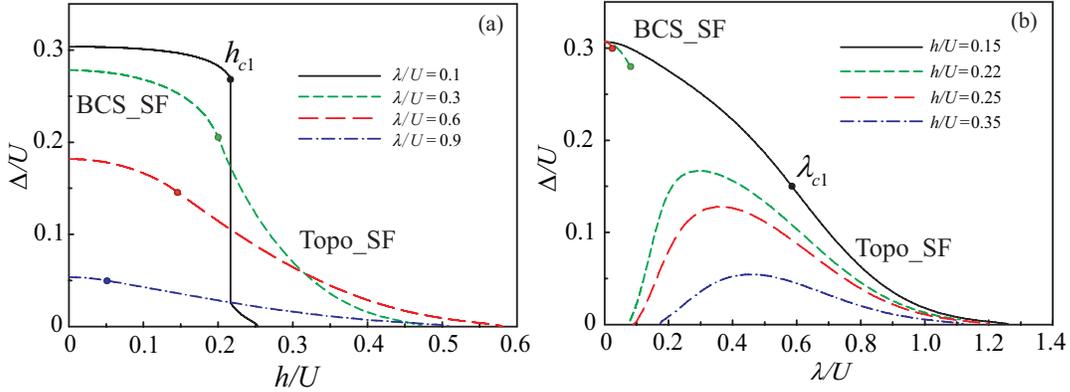}
  \vspace{-3mm}
  \caption{(a) Pairing gap $\Delta$ as a function of the Zeeman field strength $h$ for the Rashba SOC strength $\lambda/U=0.1$ (black solid line), $\lambda/U=0.3$ (green dashed line), $\lambda/U=0.6$ (red long dashed line), and $\lambda/U=0.9$ (blue dash-dotted line). (b) $\Delta$ vs $\lambda$ for  $h/U=0.15$ (black solid line), $h/U=0.22$ (green dashed line), $h/U=0.25$ (red long dashed line), and $h/U=0.35$ (blue dash-dotted line)  with $t/U=0.3$. The topological phase transition points are marked by the circle symbols.}
  \vspace{-3mm}
  \label{fig2} 
\end{figure}

To show this more clearly, we have calculated the $h$ and $\lambda$ dependences of the pairing gap, and the results of  $\Delta$ as a function of $h$ for the SOC strength $\lambda/U=0.1$ (black solid line), $\lambda/U=0.3$ (green dashed line), $\lambda/U=0.6$ (red long dashed line), and $\lambda/U=0.9$ (blue dash-dotted line) are plotted in figure \ref{fig2}a. And $\Delta$ as a function $\lambda$ for $h/U=0.15$ (black solid line), $h/U=0.22$ (green dashed line), $h/U=0.25$ (red long dashed line), and $h/U=0.35$ (blue dash-dotted line) are shown in  figure \ref{fig2}b.
It is shown clearly in figure \ref{fig2}a, the pairing gap $\Delta$ as a function $h$ has a jump around the topological phase transition point in the low $\lambda$ region, which is the characteristic of a first-order phase transition.  However, in the large $\lambda$ region, the $h$ dependence of $\Delta$  becomes a smooth decreasing behavior, which is characterized by a second-order phase transition. The topological phase transition points are marked by the circle symbols. In figure \ref{fig2}b, when $h/U=0.15$, $\Delta$ decreases smoothly with the  increase of $\lambda$. Then when $h/U=0.22$ and $0.25$, the curve of $\Delta$ vs $\lambda$  are divided into two parts: the BCS superfluid part in the low $\lambda$ region and the topological superfluid part in the larger $\lambda$ region. The pairing gap $\Delta$ in the topological superfluid increases first and then decreases as the $\lambda$ increases. However, the pairing gap in the Fermi gas is enhanced with the increase of  the SOC strength \cite{Lee17,Hu11,He12,Zhou12}. Therefore, obvious differences about the SOC effect between optical lattice and gas appear.
And when $h/U=0.35$, the BCS superfluid disappears, leaving only the topological superfluid region.
 \section{Band structure and ARPES}
Phase transition is often accompanied by the change of excitation spectra which can be obtained by the angle-resolved photoemission spectroscopy (ARPES). In the Fermi atomic system, the ARPES had been used to investigate the single-particle excitations \cite{Peter20,Stewart08,Feld2011}. In particular,
the single-particle excitations on different regions of the Brillouin zone can be measured experimentally by  ARPES, which reflects the information of pairing gap (magnitude and symmetry), band structure and Fermi surface.  The signal of ARPES is closely related to the spectral function $A({\bf k},\omega)$, $I({\bf k},\omega)=A({\bf k},\omega)n_{F}(\omega)$. As functions of momentum and energy, the spectral function can shed light on the dispersion, then get the information of band structure. Theoretically the spectral function can be obtained by the imaginary of  diagonal Green's function, $A({\bf k},\omega)=-2{\rm Im}G({\bf k},\omega)$. In this paper, the spectral function of 2D optical lattice with Rashba SOC is expressed as $A({\bf k},\omega)=-{\rm Im}[G_{1}({\bf k},\omega)+G_{2}({\bf k},\omega)]$. By using equations \ref{Greenfunction}, then the spectral function has a form as,
 \begin{eqnarray}\label{elecspectra}
A({\bf k},\omega)=\pi\sum_{a}[(U'^2_{a{\bf k}}+U^2_{a{\bf k}})\delta(\omega-E_{a{\bf k}})+(V'^2_{a{\bf k}}+V^2_{a{\bf k}})\delta(\omega+E_{a{\bf k}})],
\end{eqnarray}
A sum rule is obtained by integrating over all frequencies, $1=\int^{\infty}_{-\infty}A({\bf k},\omega)d\omega$.

\begin{figure}[ht]
  \centering
  \includegraphics[width=0.8\textwidth]{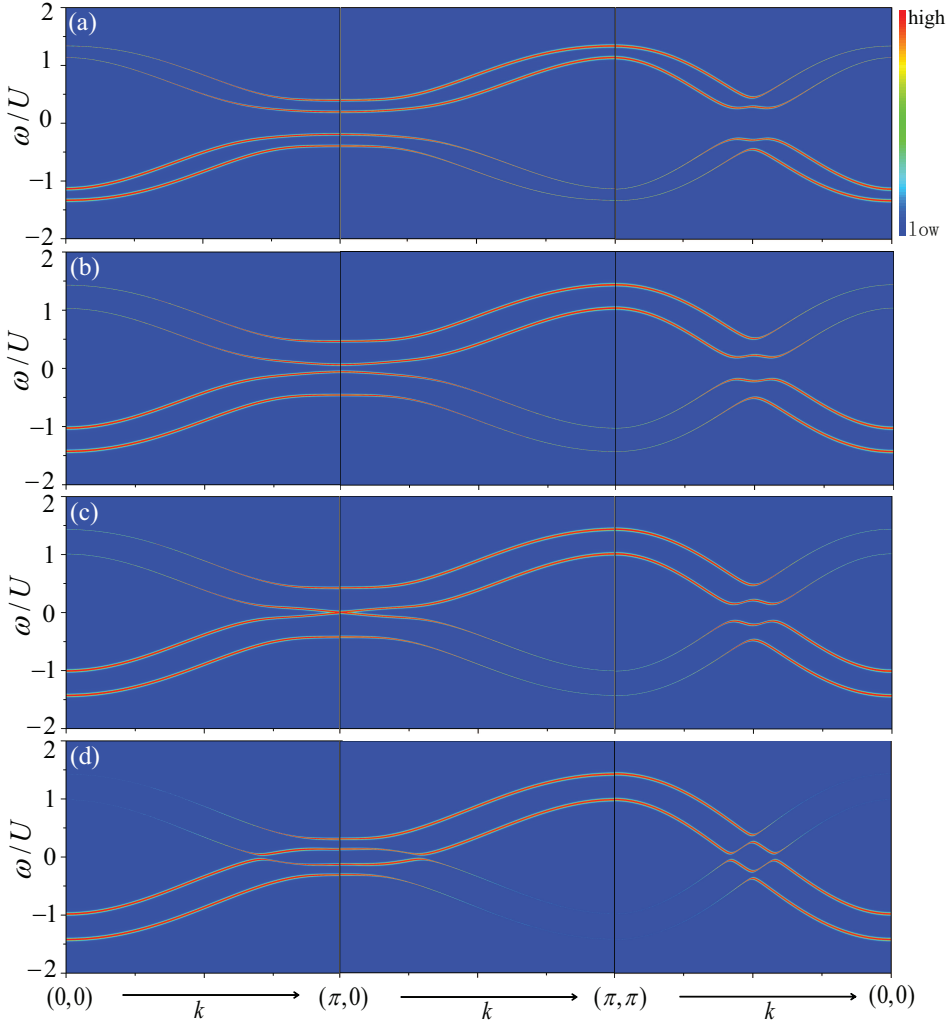}
  \vspace{-3mm}
  \caption{Dispersion along the high symmetry directions for  the Zeeman field strength (a) $h/U=0.1$ (BCS-SF),  (b) $h/U=0.2$ (BCS-SF), (c) $h_{c1}/U=0.211$ (topological transition point), and (d) $h/U=0.22$ (Topo-SF) with parameter $\lambda/U=0.15$.}
  \vspace{-3mm}
  \label{fig3} 
\end{figure}

In figure \ref{fig3}, the dispersion along the high symmetry directions in the Brillouin zone by calculating  $A({\bf k},\omega)$ are mapped for the Zeeman field strength (a) $h/U=0.1$ (BCS-SF), (b) $h/U=0.2$ (BCS-SF), (c) $h_{c1}/U=0.211$ (topological transition point), and (d) $h/U=0.22$ (Topo-SF) with parameter $\lambda/U=0.15$.
It is clearly shown that the energy spectra have four branches: $E_{1\bf{k}}$, $E_{2\bf{k}}$, $-E_{2\bf{k}}$, and $-E_{1\bf{k}}$ (from top to bottom).  With $h$ increases, the dispersion has a significant change, especially for flat band around ${\bf k_{M}}$ (here ${\bf k_{M}}=[\pi,0]$) when  $h=h_{c1}$. When $h<h_{c1}$, a excitation gap between $E_{2\bf{k}}$ and $-E_{2\bf{k}}$ exists at ${\bf k_{M}}$  and decreases as $h$ increases, and it closes ($E_{2{{\bf k}={\bf k_{M}}}}=0$) at $h=h_{c1}$ which leads to the appearance of gapless momentum points  in the Brillouin zone. Obviously a Dirac-type dispersion structure appears at $h=h_{c1}$ around ${\bf k}={\bf k_{M}}$.   In particular, stronger the Zeeman field ($h>h_{c1}$) will make the band gap to reopen. Open-close-reopen about the excitation gap reflects the phase transition from the BCS superfluid to topological superfluid.

\begin{figure}[ht]
  \centering
  \includegraphics[width=0.8\textwidth]{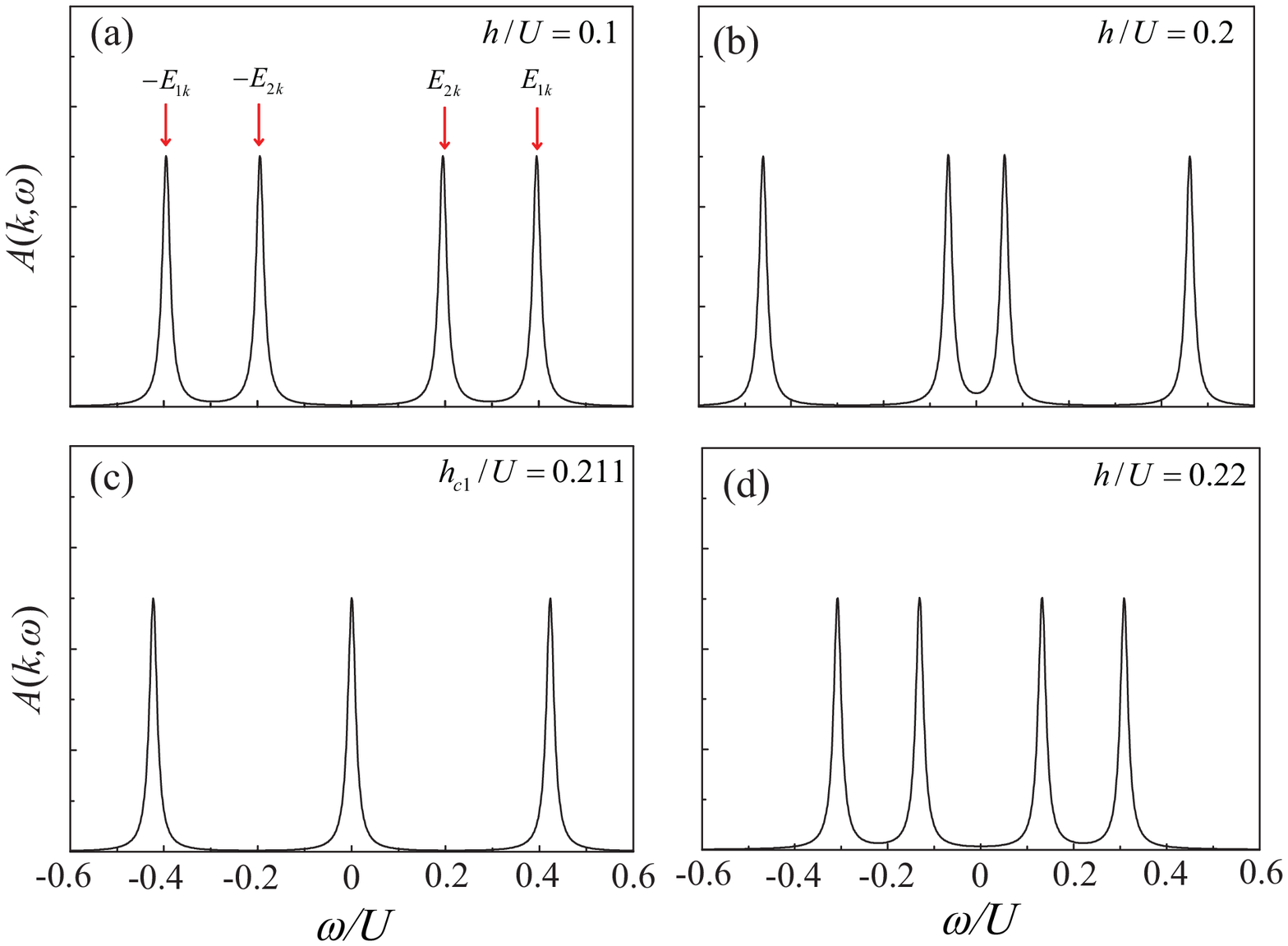}
  \vspace{-3mm}
  \caption{Spectral function $A({\bf k},\omega)$ as a function of $\omega$ at ${\bf k}={\bf k_{M}}$ for (a) $h/U=0.1$ (BCS-SF),  (b) $h/U=0.2$ (BCS-SF), (c) $h_{c1}/U=0.211$ (topological transition point), and (d) $h/U=0.22$ (Topo-SF), respectively. }
  \vspace{-3mm}
  \label{fig4} 
\end{figure}

To show the open-close-reopen of the excitation gap clearly, the SOC coupling strength dependence of spectral function $A({\bf k},\omega)$ at ${\bf k}={\bf k_{M}}$ has been calculated. Related results of $A({\bf k}={\bf k_{M}},\omega)$ as a function of ${\omega}$ for (a) $h/U=0.1$ (BCS-SF),  (b) $h/U=0.2$ (BCS-SF), (c) $h_{c1}/U=0.211$ (topological transition point), and (d) $h/U=0.22$ (Topo-SF)  are plotted in figure \ref{fig4}.
Four sharp peaks of spectral function in figure \ref{fig4}a are responding to the four branches of quasiparticle energy spectra, $-E_{1\bf{k}}$, $-E_{2\bf{k}}$, $E_{2\bf{k}}$, and $E_{1\bf{k}}$ (from left to right). With the increase of $h$, the distance of two peaks between $-E_{2\bf{k}}$ and $E_{2\bf{k}}$ reflects the magnitude of the excitation gap decreases in the BCS superfluid, and disappears at the topological transition point $h_{c1}/U=0.211$, then it reopens at the larger $h$. Moreover, from equations (\ref{Greenfunction}), the magnitude of four quasiparticle peaks (from left to right) are determined by the weight factors $V'^2_{1{\bf k}}+V^2_{1{\bf k}}$, $V'^2_{2{\bf k}}+V^2_{2{\bf k}}$, $U'^2_{2{\bf k}}+U^2_{2{\bf k}}$, $U'^2_{1{\bf k}}+U^2_{1{\bf k}}$, respectively.  At ${\bf k}={\bf k_{M}}$, the weight factors of four quasiparticle peaks are the same, which leads to the same magnitude of these peaks.

\section{Density of states (DOS)}
\begin{figure}[ht]
  \centering
  \includegraphics[width=1.0\textwidth]{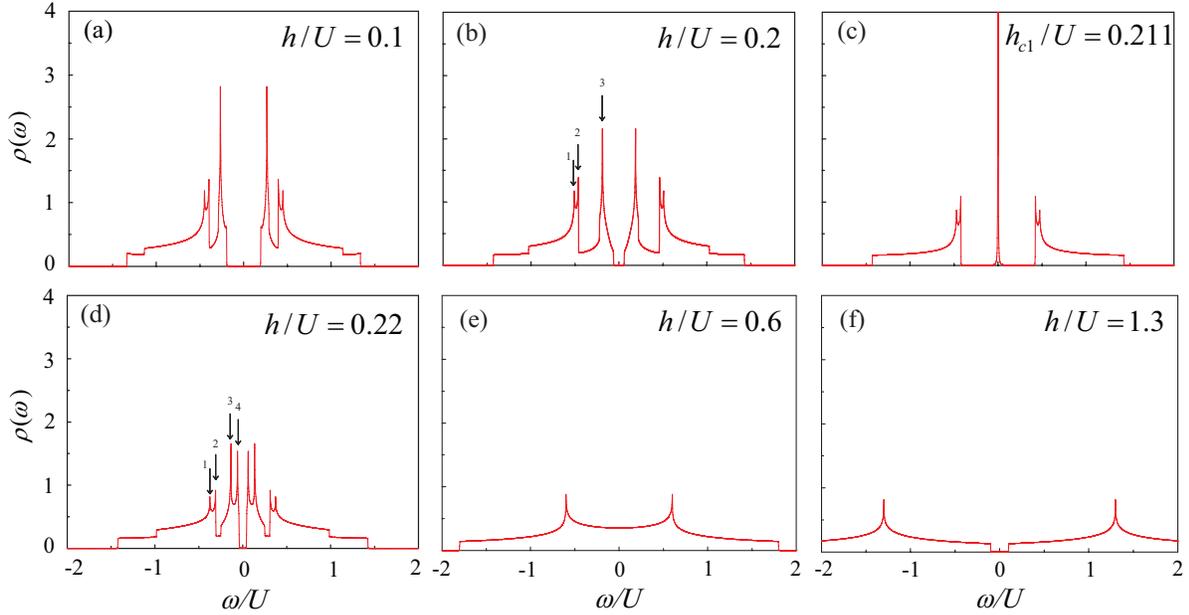}
  \vspace{-3mm}
  \caption{Density of states as a function of ${\omega}$ for (a) $h/U=0.1$ (BCS-SF),  (b) $h/U=0.2$ (BCS-SF), (c) $h_{c1}/U=0.211$ (topological transition point), (d) $h/U=0.22$ (Topo-SF), (e) $h/U=0.6$ (normal metal), and (f) $h/U=1.3$ (insulator) with $\lambda/U=0.15$.}
  \vspace{-3mm}
  \label{fig5} 
\end{figure}

Now we discuss density of states (DOS) during the topological phase transition. The density states can be obtained by spectral function, $\rho(\omega)=1/2\pi\sum_{{\bf k}}A({\bf k},\omega)$, then it has a simple form by using the equation ({\ref{elecspectra}}),
 \begin{eqnarray}\label{densitystate}
\rho(\omega)=\frac{1}{2}\sum_{{\bf k}a}[(U'^2_{a{\bf k}}+U^2_{a{\bf k}})\delta(\omega-E_{a{\bf k}})+(V'^2_{a{\bf k}}+V^2_{a{\bf k}})\delta(\omega+E_{a{\bf k}})],
\end{eqnarray}
 We calculate the energy dependence of $\rho({\omega})$ under different $h$. We choose six typical Zeeman strength parameters, (a) $h/U=0.1$ (BCS-SF),  (b) $h/U=0.2$ (BCS-SF), (c) $h_{c1}/U=0.211$ (topological transition point), (d) $h/U=0.22$ (Topo-SF), (e) $h/U=0.6$ (normal metal), and (f) $h/U=1.3$ (insulator), and plot $\rho({\omega})$ in figure \ref{fig5}.
It is shown that the DOS is strongly dependent on $h$, and is symmetric between positive and negative energy. Due to the modulation of SOC on the DOS, the DOS is affected not only by the pairing gap, but also by the strength of SOC.  At the BCS superfluid, the weight of DOS disappears at the Fermi energy ($\omega=0$) and locates at the high-binding energy region. With the increase of $h$, the weight of DOS moves towards to the Fermi energy, and shows a sharp peak at the Fermi energy when $h=h_{c1}$, in which the excitation gap of $E_{2{\bf k_{M}}}$ closes at $k_{M}$, ($E_{2{\bf k_{M}}}=0$). When $h>h_{c1}$, the weight of DOS moves back to the high-binding energy from the Fermi energy as the excitation gap reopens in the topological superfluid. Moreover, stronger $h$ in figure \ref{fig5}e will destroy the superfluid (then $\Delta=0$) and the normal metal phase appears, thus the weight of DOS at the Fermi energy appears. When $h$ is over $h_{c3}$, the system will enter the insulator phase, in which the weight of DOS at the Fermi energy disappears as a result of the band gap. In particular, when the system enters the topological superfluid from the BCS superfluid, an interesting phenomenon about DOS can be found, i.e., there are three characteristic peaks in $h/U=0.2$ (BCS-SF) while four characteristic peaks appear in $h/U=0.22$ (Topo-SF). The characteristic peaks are marked by the arrows with a number.  The abrupt change of characteristic peaks in DOS during the topological phase transition indicates the change of Van Hove singularities which is associated with the quasiparticle energy spectra, $\rho({\omega})\propto (\nabla_{\bf k} E_{a{\bf k}})^{-1}$. Therefore, to study this problem clearly, we discuss the quasiparticle energy spectra from the BCS superfluid to the topological superfluid in the Brillouin zone.

\begin{figure}[ht]
  \centering
  \includegraphics[width=0.8\textwidth]{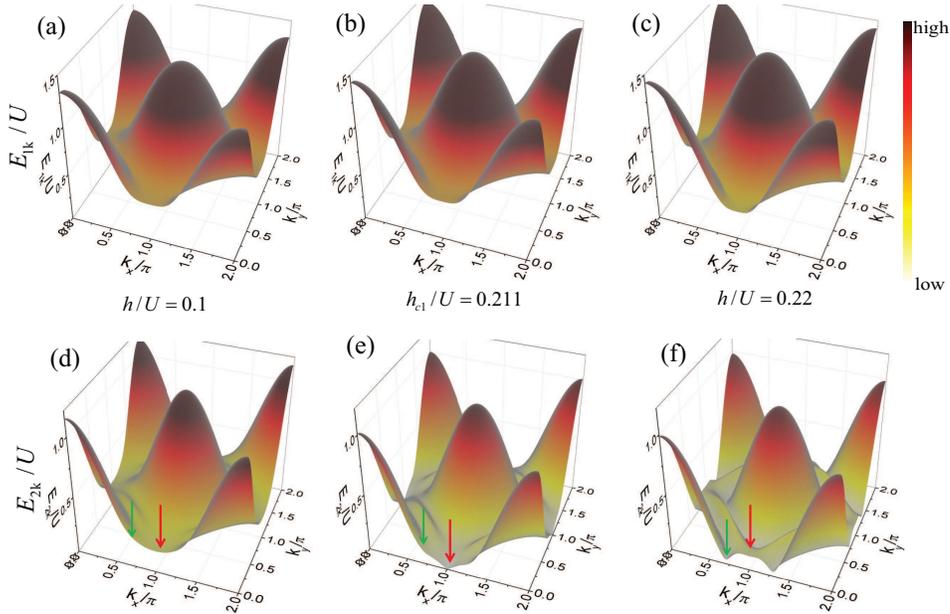}
  \vspace{-3mm}
  \caption{Quasiparticle energy spectra $E_{1\bf{k}}$ (top row) and $E_{2\bf{k}}$ (bottom row)  for  (a)(d) $h/U=0.1$, (b)(e) $h_{c1}/U=0.211$, (c)(f) $h/U=0.22$ from the BCS superfluid to the topological superfluid in the full Brillouin zone.}
  \vspace{-3mm}
  \label{fig6} 
\end{figure}

\begin{figure}[ht]
  \centering
  \includegraphics[width=0.9\textwidth]{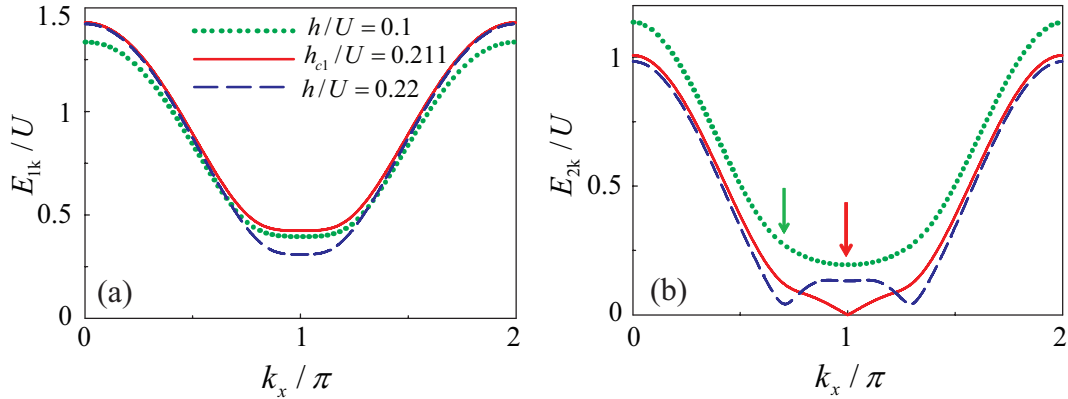}
  \vspace{-3mm}
  \caption{ Quasiparticle energy spectra (a) $E_{1{\bf k}}$ and (b) $E_{2{\bf k}}$ along the high symmetry direction $[0,0]\rightarrow [\pi,0]\rightarrow [2\pi,0]$ for $h/U=0.1$ (green dotted line), $h_{c1}/U=0.211$ (red solid line), and $h/U=0.22$ (blue dashed line) with $\lambda/U=0.15$.}
  \vspace{-3mm}
  \label{fig7} 
\end{figure}

Now we plot $E_{1\bf{k}}$ (top row) and $E_{2\bf{k}}$ (bottom row) for (a)(d) $h/U=0.1$, (b)(e) $h_{c1}/U=0.211$, (c)(f) $h/U=0.22$  in figure {\ref{fig6}}.
Our results show that with the increase of $h$, the low-energy quasiparticle spectrum $E_{2\bf{k}}$ changes obviously from the BCS superfluid to the topological superfluid. On the one hand, around ${\bf k_{M}}=[\pi,0]$ marked by the red arrow, an excitation gap opens in both the BCS superfluid and the topological superfluid, and it vanishes at the topological phase transition point in figure \ref{fig6}e indicating the appearance of the gapless excitations. On the other hand, along the high symmetry direction $[0,0]\rightarrow [\pi,0]\rightarrow [2\pi,0]$,  the lowest point of $E_{2\bf{k}}$ is located at ${\bf k_{M}}$ marked by the red arrow in the BCS superfluid, while in the topological superfluid state, it is located at position marked by the green arrow. To see these points clearly, the momentum dependence of $E_{1\bf{k}}$ and $E_{2\bf{k}}$  was calculated along $[0,0]\rightarrow [\pi,0]\rightarrow [2\pi,0]$ in the Brillouin zone. The related results obtained for (a) $E_{1\bf{k}}$  and (b) $E_{2\bf{k}}$  as a function of momentum at $\lambda/U=0.15$ for $h/U=0.1$ (green dotted line), $h_{c1}/U=0.211$ (red solid line), and $h/U=0.22$ (blue dashed line) are plotted in figure \ref{fig7}.
 In the BCS superfluid with a small $h$, the lowest point for both $E_{1{\bf k}}$  and $E_{2{\bf k}}$  appears at ${\bf k_{M}}$. When $h=h_{c1}$, $E_{2{\bf k}}$ has a gapless excitation at ${\bf k_{M}}$ and  a linear Dirac-type dispersion appears. Going on increasing of $h$, the topological superfluid appears, in which the lowest point of $E_{2\bf{k}}$  shifts from $\pi$ to a smaller momentum. Therefore, the topology of the quasiparticle energy spectrum $E_{2{\bf k}}$ has a dramatic change during the topological phase transition from the BCS superfluid to the topological superfluid. Obviously, there are more momentum points satisfying $\nabla_{\bf k} E_{2{\bf k}}=0$ in the topological superfluid. From $\rho({\omega})\propto (\nabla_{\bf k} E_{a{\bf k}})^{-1}$ as mentioned above, the DOS of the topological superfluid will have more peak. Moreover, the topology of the low-energy quasiparticle energy spectrum in 1D SOC Fermi gas has a similar change with increase of the Zeeman field \cite{Fan2022}.

\begin{figure}[ht]
  \centering
  \includegraphics[width=0.9\textwidth]{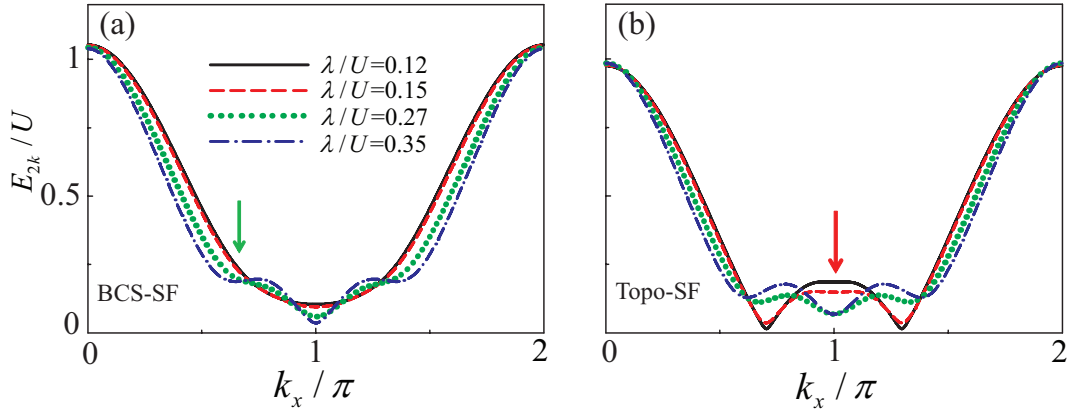}
  \vspace{-3mm}
  \caption{Along $[0,0]\rightarrow [\pi,0]\rightarrow [2\pi,0]$, (a) $h/U$=0.18 (BCS-SF) and (b) $h/U=0.225$ (Topo-SF) of $E_{2{\bf k}}$ for the SOC strength $\lambda/U=0.12$ (black solid line), $\lambda/U=0.15$ (red dashed line), $\lambda/U=0.27$ (green dotted line) and $\lambda/U=0.35$ (blue dash-dotted line).}
  \vspace{-3mm}
  \label{fig8} 
\end{figure}

Moreover, from the viewpoint of the topology of the low-energy quasiparticle energy spectrum, we try to discuss the change from the first-order (low $\lambda$) phase transition to the second-order (larger $\lambda$) phase transition  with the increase of $\lambda$ as mentioned in figure \ref{fig2}. Along $[0,0]\rightarrow [\pi,0]\rightarrow [2\pi,0]$ in the Brillouin zone, we plot $E_{2{\bf k}}$ at (a) $h/U$=0.18 (BCS-SF) and (b) $h/U=0.225$ (Topo-SF) for the SOC strength $\lambda/U=0.12$ (black solid line), $\lambda/U=0.15$ (red dashed line), $\lambda/U=0.27$ (green dotted line) and $\lambda/U=0.35$ (blue dash-dotted line) in figure \ref{fig8}.
It is shown that the topology of $E_{2{\bf k}}$ in both the BCS superfluid and topological superfluid is strongly dependent on the SOC strength. First, when $\lambda=0$ and $h>0$ (without SOC), a like traditional BCS state of supeconductor will keep un-magnetized until a critical Zeeman field. Then a first-order phase transition appears from the BCS superfluid to the unpaired magnetized metal state \cite{Sheehy2007}. Second, for the BCS superfluid, as $\lambda$ increasing, a kink structure around the green arrow appear. Third, for the topological superfluid, $E_{2{\bf k}}$ around $[\pi,0]$ (marked by the red arrow) is flat in the low $\lambda$ region and it decreases as $\lambda$ increasing. These results indicate that the symmetry of $E_{2{\bf k}}$  both the BCS superfluid and topological superfluid is changed with the increase of $\lambda$. We find that the symmetry change of energy spectra is accompanied by the change from first-order phase transition to second-order phase transition. In a word, the phase transition from the first-order to the second-order phase transition with the increase of $\lambda$ may be related to the symmetry change of energy spectra.
\section{Doping dependence of phase diagram}
\begin{figure}[ht]
  \centering
  \includegraphics[width=0.8\textwidth]{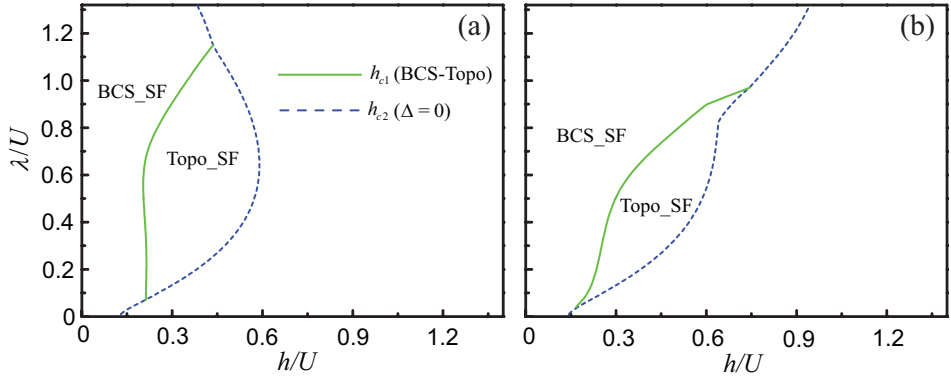}
  \vspace{-3mm}
  \caption{Doping dependence of phase diagram for different average occupancy numbers: (a) $n=0.9$ and (b) $n=0.8$.}
  \vspace{-3mm}
  \label{fig9} 
\end{figure}

For the Fermi system, the average occupancy number $n=1-\delta$, where $\delta$ is the doping concentration, and it indicates how far away the system is from half-filling. Doping can change the concentration of carrier, then greatly influence phase diagram and related physical properties.  Now, we will discuss the phase diagram of  2D Rashba SOC lattice system when the system is away from half-filling.  In figure \ref{fig9}, we plot the phase diagram in the $h-\lambda$ plane for different average occupancy numbers: (a) $n=0.9$ and (b) $n=0.8$ at $t/U=0.3$.
The calculated results reveal that the phase diagram is strongly dependent on doping.  As $n$ decreases from half-filling, the topological superfluid region has shrunk while the region of BCS superfluid enlarges. Therefore, it is harder to realize the topological superfluid when the system is away from half-filling. Moreover, the insulator region disappears, which can be explained by the change of the quasiparticle energy spectra relative to the Fermi energy as $n$ decreases: the low-energy  quasiparticle energy spectrum always pass through the Fermi energy when the doping concentration exceeds a small amount. This phenomenon had been discussed in 1D optical lattice with Raman SOC \cite{Han2022}.

\section{Summary }
The phase diagram, band structure and density of states (DOS) of Fermi atomic system in 2D optical lattice with Rashba spin-orbit coupling are discussed within the mean-field theory. The phase diagram in the $h-\lambda$ plane shows that the topological superfluid appears under proper Zeeman field strength and SOC strength, and has an optimal Rashba SOC strength where the topological superfluid exists in a wide range of Zeeman field and the pairing gap is still large.   Moreover, by calculating the spectral function, we obtain the change of both band structure and DOS during the phase transition from BCS superfluid to topological superfluid. At the topological phase transition point, a Dirac dispersion appears in the low-energy quasiparticle energy spectrum, which leads to a sharp peak in the DOS at the Fermi energy. In the topological superfluid, the characteristic peaks of the DOS are one more than that of the BCS superfluid, which can be attributed to  the topology change of the low-energy quasiparticle energy spectrum during the topological phase transition.
\section*{Acknowledgments}
We are grateful to fruitful discussions with Professor Peng Zou.
This work was supported by the funds from the National
Natural Science Foundation of China under Grant Nos.11547034, 11804177.


\section*{References}

\begin{thebibliography}{0}
\bibitem{Zhai15} Zhai H 2015 {\it Rep. Prog. Phys.} {\bf 78} 026001
\bibitem{Jayantha11}Vyasanakere J P and Shenoy V B 2011 {\it Phys. Rev. B} {\bf 83} 094515
\bibitem{Jayantha11-2} Vyasanakere J P, Zhang S Z and Shenoy V B 2011 {\it Phys. Rev. B} {\bf 84} 014512
\bibitem{Hu2011}Hu H, Jiang L, Liu X-J and Pu H 2011 {\it Phys. Rev. Lett.} {\bf 107} 195304
\bibitem{He2013} He L Y and Huang X-G 2013 {\it Ann. Phys.} {\bf 337} 163-207
\bibitem{Vyasanakere12} Vyasanakere J P and Shenoy V B 2012 {\it Phys. Rev. A} {\bf 86} 053617
\bibitem{Zhang2013} Zhang S-S, Yu X-L, Ye J W and Liu W M 2013 {\it Phys. Rev. A} {\bf 87} 063623

\bibitem{Zhou11} Zhou J, Zhang W and Yi W 2011  {\it Phys. Rev. A} {\bf 84} 063603
\bibitem{Liao10} Liao Y, Rittner A S C, Paprotta T, Li W, Partridge G B, Hulet R G, Baur S K and Mueller E J 2010  {\it Nature} (London) {\bf 467} 567
\bibitem{Cao14} Cao Y, Zou S-H, Liu X-J, Yi S, Long G-L and Hu H 2014  {\it Phys. Rev. Lett.} {\bf 113} 115302
\bibitem{Zhou14} Zhou L H, Cui X L and Yi W 2014 {\it Phys. Rev. Lett.} {\bf 112} 195301
\bibitem{Hu13} Hu H and Liu X-J 2013 {\it New J. Phys.} {\bf 15} 093037
\bibitem{Alicea2012} Alicea J 2012 {\it Rep. Prog. Phys.} {\bf 75} 076501
\bibitem{Lin2009} Lin Y-J, Compton R L, Jim\'enez-Garc\'ia K, Proto J V and Spielman I B  2009 {\it Nature} {\bf 462} 628
\bibitem{Lin2011}  Lin Y-J, Compton R L, Jim\'enez-Garc\'ia K, Phillips W D, Proto J V and Spielman I B 2011 {\it Nat. Phys.} {\bf 7} 531
\bibitem{Lin2011-2} Lin Y-J, Jim\'enez-Garc\'ia K,  Spielman I B 2011 {\it Nature} (London) {\bf 471} 83
\bibitem{Galitski13} Galitski V and Spielman I B, 2013 {\it Nature} {\bf 494} 49-54
\bibitem{Cheuk12} Cheuk L W, Sommer A T, Hadzibabic Z, Yefsah T, Bakr W S and Zwierlein M W 2012 {\it Phys. Rev. Lett.} {\bf 109} 095302
\bibitem{Wang12} Wang P, Yu Z-Q, Fu Z, Miao J, Huang L, Chai S, Zhai H and Zhang J 2012 {\it Phys. Rev. Lett.} {\bf 109} 095301
\bibitem{Zhang14} Zhang J, Hu H, Liu X-J and Pu H 2014 {\it Annual Review of Cold Atoms and Molecules} {\bf 2} 81
\bibitem{Williams13} Williams R A, Beeler M C, LeBlanc L J, Jim\'enez-Garc\'ia K and Spielman I B 2013 {\it Phys. Rev. Lett.} {\bf 111} 095301
\bibitem{Meng16} Meng Z M, Huang L H, Peng P, Li D H, Chen L C, Xu Y, Zhang C W, Wang P J and Zhang J 2016 {\it Phys. Rev. Lett.} {\bf 117} 235304
\bibitem{Huang18} Huang L H, Peng P, Li D H, Meng Z M, Chen L C, Qu C L, Wang P J, Zhang C W and Zhang J 2018 {\it Phys. Rev. A} {\bf 98} 013615
\bibitem{Burdick16} Burdick N Q, Tang Y J and Lev B L 2016 {\it Phys. Rev. X} {\bf 6} 031022
\bibitem{Huang16} Huang L H, Meng Z M, Wang P J, Peng P, Zhang S-L, Chen L C, Li D H, Zhou Q and Zhang J 2016 {\it Nat. Phys.} {\bf 12} 540
\bibitem{Wu2016} Wu Z, Zhang L, Sun W, Xu X-T, Wang B Z, Ji S-C, Deng Y J, Chen S, Liu X-J and Pan J-W  2016 {\it Science} {\bf 354} 6308
\bibitem{Zhang18} Zhang S C, He C D, Hajiyev E, Ren Z J, Song B and Jo G-B 2018 {\it Scientific Reports} {\bf 8} 18005
\bibitem{Yang2012} Yang X S and Wan S L 2012 {\it Phys. Rev. A} {\bf 85} 023633
\bibitem{Liu2012} Liu X-J, Jiang L, Pu H and Hu H 2012 {\it Phys. Rev. A} {\bf 85} 021603(R)
\bibitem{Sheehy2007}Sheehy D E and Radzihovsky L 2007 {\it Ann. Phys.} {\bf 322} 1790
\bibitem{Zhang13} Zhang W and Yi W 2013 {\it Nat Commun} {\bf 4} 2711
\bibitem{Bloch08} Bloch I, Dalibard J and Zwerger W 2008 {\it Rev. Mod. Phys.} {\bf 80} 885
\bibitem{Mitra18} Mitra D, Brown P T, Guardado-Sanchez E, Kondov S S, Devakul T, Huse D A, Schau\ss{} P and Bakr W S 2018 {\it Nat. Phys.}  {\bf 14} 173
\bibitem{Hackermuller10} Hackerm{\"u}ller L, Schneider U, Moreno-Cardoner M, Kitagawa T, Best T, Will S, Demler E, Altman E, Bloch I and Paredes B 2010 {\it Science} {\bf 327} 1621
\bibitem{Peter20} Brown P T, Guardado-Sanchez E, Spar B M, Huang E W, Devereaux T P and Bakr W S 2020 {\it Nat. Phys.} {\bf 16} 26-31

\bibitem{Scalettar89} Scalettar R T, Loh E Y, Gubernatis J E, Moreo A, White S R, Scalapino D J, Sugar R L and Dagotto E 1989 {\it Phys. Rev. Lett.} {\bf 62} 1407 (1989).
\bibitem{Moreo91} Moreo A and Scalapino D J 1991  {\it Phys. Rev. Lett.}  {\bf 66} 946
\bibitem{Singer96} Singer J M, Pedersen M H, Schneider T, Beck H and  Matuttis H-G 1996  {\it Phys. Rev. B} {\bf 54} 1286
\bibitem{Paiva04} Paiva T, R. dos Santos  R, Scalettar R T and Denteneer P J H 2004 {\it Phys. Rev. B} {\bf 69} 184501
\bibitem{Ho04} Ho A F, Cazalilla M A and Giamarchi T 2009  {\it Phys. Rev. A} {\bf 79} 033620
\bibitem{Xu2014} Xu Y, Qu C L, Gong M and Zhang C W 2014 {\it Phys. Rev. A} {\bf 89} 013607
\bibitem{Koinov17} Koinov Z and Pahl S 2017 {\it Phys. Rev. A} {\bf 95} 033634
\bibitem{Sun2013} Sun Q, Zhu G B, Liu W-M and Ji A-C 2013 {\it Phys. Rev. A} {\bf 88} 063637
\bibitem{Gremaud13} Min\'a\v{r} J and Gr\'emaud B 2013 {\it Phys. Rev. B} {\bf 88} 235130
\bibitem{Riera13} Riera J A 2013 {\it Phys. Rev. B} {\bf 88} 045102
\bibitem{Goldman2009} Goldman N, Kubasiak A, Bermudez A, Gaspard P, Lewenstein M and Martin-Delgado M A 2009
{\it Phys. Rev. Lett.} {\bf 103} 035301
\bibitem{Iskin2013} Iskin M 2013 {\it Phys. Rev. A} {\bf 88} 013631
\bibitem{Burrello2013} Burrello M, Fulga I C, Alba E, Lepori L and Trombettoni A 2013 {\it Phys. Rev. A} {\bf 88} 053619
\bibitem{Wang2013} Wang L and Fu L B 2013 {\it Phys. Rev. A} {\bf 87} 053612
\bibitem{Qu2013} Qu C L, Zheng Z, Gong M, Xu Y, Mao L, Zou X B, Gao G C and Zhang C W 2013 {\it Nat. Commun.} {\bf 4} 2710
\bibitem{Jia2019} Jia W, Huang Z-H, Wei X, Zhao Q, and Liu X-J 2019 {\it Phys. Rev. B} {\bf 99} 094520
\bibitem{Wu2021} Wu Y-J, Luo X-W, Hou J, and Zhang C 2021 {\it Phys. Rev. A} {\bf 103} 013307
\bibitem{Liu07} Liu X-J, Hu H and Drummond P D 2007 {\it Phys. Rev. A} {\bf 76} 043605
\bibitem{Wu13} Wu F, Guo G-C, Zhang W and Yi W 2013 {\it Phys. Rev. Lett.} {\bf 110} 110401
\bibitem{Liu2013} Liu X-J and Hu H 2013 {\it Phys. Rev. A} {\bf 88} 023622
\bibitem{Yuan2021} Yuan N F Q and Fu L 2021 {\it PNAS} {\bf 118} e2019063118
\bibitem{Wang2018} Wang B-Z, Lu Y-H, Sun W, Chen S, Deng Y J, and Liu X-J 2018 {\it Phys. Rev. A} {\bf 97} 011605(R)
\bibitem{Juzeliunas10} Juzeli\={u}nas G, Ruseckas J and Dalibard J 2010 {\it Phys. Rev. A} {\bf 81} 053403

\bibitem{Campbell11} Campbell D L, Juzeli\={u}nas G and Spielman I B 2011 {\it Phys. Rev. A} {\bf 84} 025602

\bibitem{Xu12} Xu Z F and You L 2012 {\it Phys. Rev. A} {\bf 85} 043605

\bibitem{Xu13} Xu Z F, You L and Ueda M 2013 {\it Phys. Rev. A} {\bf 87} 063634

\bibitem{Zhou19} Zhou X F, Luo X-W, Chen G, Jia S T and Zhang C W 2019 {\it Phys. Rev. A} {\bf 100} 063630

\bibitem{Sau11} Sau J D, Sensarma R, Powell S, Spielman I B and Sarma S D 2011 {\it Phys. Rev. B} {\bf 83} 140510(R)

\bibitem{Anderson13} Anderson B M, Spielman I B and Juzeli\={u}nas 2013 {\it Phys. Rev. Lett.} {\bf 111} 125301

\bibitem{Dalibard11} Dalibard J and Gerbier F 2011 {\it Rev. Mod. Phys.} {\bf 83} 1523-1543

\bibitem{Zhao20} Zhao H, Gao X, Liang W, Zou P and Yuan F 2020 {\it New J. Phys.} {\bf 22} 093012
\bibitem{Han2022} Han R, Yuan F and Zhao H 2022 {\it Europhysics Letters} {https://doi.org/10.1209/0295-5075/ac39ec}
\bibitem{Lee17} Lee J and Kim D-H 2017 {\it Phys. Rev. A} {\bf 95} 033609
\bibitem{Hu11} Hu H, Jiang L, Liu X-J and Pu H 2011 {\it Phys. Rev. Lett.} {\bf 107} 195304
\bibitem{He12} He L Y and Huang X-G 2012 {\it Phys. Rev. Lett.} {\bf 108} 145302
\bibitem{Zhou12} Zhou K Z and Zhang Z D 2012 {\it Phys. Rev. Lett.} {\bf 108} 025301
\bibitem{Stewart08} Stewart J T, Gaebler J P and Jin D S 2008 {\it Nature} {\bf 454} 744
\bibitem{Feld2011} Feld M, Fr{\"o}hlich B, Vogt E, Koschorreck M and K{\"o}hl  M 2011  {\it Nature} {\bf 480} 75
\bibitem{Fan2022} Fan G, Chen X L and Zou P  2022  {\it Front. Phys.}  {\bf 17} 52502

\end{thebibliography}
\end{document}